\documentclass[letterpaper,12pt]{article}

\pdfoutput=1
\usepackage{amsmath}
\usepackage{hyperref}
\usepackage{bm}

\makeatletter
\renewcommand\section{\@startsection {section}{1}{\z@}%
{-3.5ex \@plus -1ex \@minus -0.2ex}%
{2.3ex \@plus 0.2ex}%
{\normalfont\normalsize\bfseries}}

\renewcommand\subsection{\@startsection{subsection}{2}{\z@}%
{-3.25ex \@plus -1ex \@minus -0.2ex}%
{1.5ex \@plus 0.2ex}%
{\normalfont\normalsize\bfseries}}

\def\@seccntformat#1{\csname the#1\endcsname.\quad}
\makeatother

\newcommand{\oline}[1]{\overline{\mkern-1.0mu#1\mkern0.0mu}}
\newcommand{\xoline}[1]{\hspace{0.30em}\overline{\mkern-5.0mu#1\mkern-2.0mu}\hspace{0.12em}}
\newcommand{\uline}[1]{\underline{\mkern0.0mu#1\mkern-1.0mu}}
\newcommand\redots{\makebox[0.85em][c]{.\hfil.\hfil.}}

\begin{document}

\setlength{\baselineskip}{3.75ex}

\noindent
\textbf{\LARGE Physical, subjective and analogical}\\[2ex]
\textbf{\LARGE probability}\\[3ex]

\noindent
\textbf{Russell J. Bowater}\\
\emph{Independent researcher,
Doblado 110, Col.\ Centro, City of Oaxaca, C.P.\ 68000, Mexico.\\
Email address: as given on arXiv.org. Twitter profile:
\href{https://twitter.com/naked_statist}{@naked\_statist}\\ Personal website:
\href{https://sites.google.com/site/bowaterfospage}{sites.google.com/site/bowaterfospage}}
\\[2ex]

\noindent
\textbf{\small Abstract:}
{\small
The aim of this paper is to show that the concept of probability is best understood by dividing
this concept into two different types of probability, namely physical probability and analogical
probability.
Loosely speaking, a physical probability is a probability that applies to the outcomes of an
experiment that have been judged as being equally likely on the basis of physical symmetry.
Physical probabilities are arguably in some sense `objective' and possess all the standard
properties of the concept of probability.
On the other hand, an analogical probability is defined by making an analogy between the
uncertainty surrounding an event of interest and the uncertainty surrounding an event that has a
physical probability.
Analogical probabilities are undeniably subjective probabilities and are not obliged to have all
the standard mathematical properties possessed by physical probabilities, e.g.\ they may not have
the property of additivity or obey the standard definition of conditional probability.
Nevertheless, analogical probabilities have extra properties, which are not possessed by physical
probabilities, that assist in their direct elicitation, general derivation, comparison and
justification. More specifically, these properties facilitate the application of analogical
probability to real-world problems that can not be adequately resolved by using only physical
probability, e.g.\ probabilistic inference about hypotheses on the basis of observed data.
Careful definitions are given of the concepts that are introduced and, where appropriate, examples
of the application of these concepts are presented for additional clarity.}
\\[3ex]

\pagebreak
\section{Introduction}

While the study of probability in the field of mathematics is highly developed, it has proved, over
the years, to be difficult to find an adequate answer to the simple question of what is the
philosophical meaning of the concept of probability, see for example Fine~(1973), Gillies~(2000)
and Eagle~(2011). Nevertheless, resolving this issue may have substantial implications in terms of
how probabilistic methods are applied to tackle real-world problems, e.g.\ the problem of how
statistical inference should be performed in any given situation.

With regard to this issue, two different types of probability will be identified in the present
paper. The first type of probability will be called physical probability. A physical probability
will be defined in Section~\ref{sec1} but, loosely speaking, it is a probability that applies to
the outcomes of an experiment that have been judged as being equally likely on the basis of
physical symmetry.
The second type of probability will be called analogical probability. An analogical probability is
defined by making an analogy between the uncertainty surrounding an event of interest and the
uncertainty surrounding an event that has a physical probability.

A physical probability will be considered as being the type of probability that is the closest we
can get to an objective probability, while an analogical probability will be classed, without
doubt, as being a subjective probability. Apart from these two types of probability, no space will
be allowed for other types of probability, which means that a proposed probability of another type
that can not be dismissed as being a flawed definition of probability will be regarded as being a
way of trying to measure or estimate either physical or analogical probability.

We will define physical probability, or in other words `objective' probability, such that this type
of probability possesses all of the standard properties of the concept of probability. For example:

\vspace{2ex}
\noindent
\textbf{Some standard properties of probability}

\vspace{1ex}
\noindent
1) A probability must lie in the interval $[0,1]$.\\
2) If the probability $P(E)$ of an event $E$ is zero then $E$ is impossible, while if $P(E)=1$ then
$E$ is certain. On the other hand, if $0<P(E)<1$ then $E$ may or may not occur.\\
3) If $A$ and $B$ are mutually exclusive events then $P(A \cup B) = P(A) + P(B)$, i.e.\
probabilities are additive.\\
4) The probability of an event $B$ conditional on an event $A$ having occurred, i.e.\ the
probability $P(B\,|\,A)$, is defined by the expression: $P(B\,|\,A) = P(A \cap B) / P(A)$.

\vspace{2ex}
\noindent
We will avoid the common practice of referring to the kinds of \emph{properties} of physical
probabilities just listed as being \emph{rules} or \emph{axioms} since these are arguably
inappropriate terms to use which can lead to confusion.

Analogical probabilities, i.e.\ subjective probabilities, do not need to have any of the properties
of physical probabilities although usually, for convenience, we will restrict analogical
probabilities so that they do have one or more of these properties, e.g.\ it is usual for
subjective probabilities to be restricted so that they at least have properties (1) and (2) just
listed. Nevertheless, many authors have tried to argue that it is unacceptable for subjective
probabilities not to have all the mathematical properties that are possessed by physical
probabilities. These types of argument fall broadly into two categories:

\vspace{2ex}
\noindent
1) Arguments based on proposing a set of axioms that we may reasonably expect would naturally be
adhered to by any rational agent and then showing that if the agent always adheres to these axioms,
then he will always follow the standard `rules' of probability. Examples of arguments of this type
can be found, for example, in Savage~(1954), Fishburn~(1986), Bernardo and Smith~(1994) and
Jaynes~(2003).

\vspace{2ex}
\noindent
2) Arguments based on showing that if an individual chooses not to adhere to the stan\-dard `rules'
of probability, then the individual will suffer undesirable consequences such as a guaranteed
financial loss. Dutch book arguments clearly fall into this category, see for example Ramsey~(1926)
and de Finetti~(1937).

\vspace{2ex}
However, it can be easily appreciated that the most popular arguments falling into the first of
these categories are individually based on one or more axioms that are not so reasonable, and it
would seem fair to expect that this would be the case for all arguments that may fall in this
category.
For example, axioms that may seem to be quite acceptable in the simplest of examples but which are
certainly not acceptable in all examples that are imaginable, or axioms that would only be
acceptable to someone who, for some unclear reason, is already sold on the idea that subjective
probabilities should obey the standard `rules' of probability.
Furthermore, it is apparent that the kind of undesirable consequences of not following the standard
`rules' of probability that are identified by the types of argument falling into the second
category just mentioned will only arise by constraining the individual in question to obey
additional rules that, from any real practical viewpoint, he is not naturally obliged to obey,
e.g.\ to buy and sell gambles at the same price.
Therefore, we will justifiably put to one side arguments that attempt to make the case that
subjective probabilities are obliged to have the same mathematical properties as physical
probabilities.
By doing this, we are, for example, clearly opening the door to the possibility that post-data
probability distributions can be placed over model parameters without using Bayes' theorem.

Finally, it will often be useful to think of the class of all analogical probabilities, i.e.\
subjective probabilities, as being divided into two sub-categories, namely personal subjective
probabilities and communal subjective probabilities.
We will define a personal subjective probability of any given event as not only being a probability
that is subjective but a probability that we would not expect to be accepted as the probability of
the event concerned by many other people apart from an individual of interest.
On the other hand, a communal subjective probability will be defined as a probability of any given
event that despite being subjective would be accepted as the probability of this event by many (if
not most) individuals who are in the same broadly defined information state.

Let us now briefly describe the structure of the paper. In the next section, the concept of
physical probability is defined. In particular, this section begins by defining the key notion of
\emph{similarity}, which underlies the definitions of both physical and analogical probability.
Separate definitions of the concept of physical probability are then given for the cases where this
type of probability is discrete and where it is continuous.

Following on, having used the notion of physical probability to define the concepts of discrete and
continuous reference sets of events in Section~\ref{sec5}, these latter concepts are called upon in
Sections~\ref{sec3} and~\ref{sec4}, along with the idea of similarity, to define the concept of
analogical probability.
In particular, the notion of non-additive analogical probability is defined and analysed in
Section~\ref{sec3}, which is an analysis that is then used to justify the definition and discussion
of the notion of additive analogical probability in Section~\ref{sec4}.
This latter section is a long section that contains various definitions that relate to how
practical issues can be resolved by applying concepts associated with analogical probability, e.g.\
probability elicitation via the concept of the internal strength of a probability distribution and
comparisons of the representativeness of already elicited or derived distributions via the concept
of the external strength of a distribution.
In Section~\ref{sec8}, this latter concept is then applied to a long-running controversy concerning
what and how big is the advantage of using the fiducial argument as opposed to Bayesian reasoning
to address a particular class of problems in statistical inference.
Of special interest to some readers may be the discussion in Section~\ref{sec7} of how the concept
of frequentist probability fits into the ideas put forward in the present paper.
The final section of the paper, i.e.\ Section~\ref{sec9}, contains some concluding remarks.

\vspace{3ex}
\section{Physical probabilities}
\label{sec1}

\vspace{0.75ex}
\noindent
\textbf{Definition 1: Similarity}

\vspace{1.25ex}
\noindent
Let $S(A,B)$ denote the similarity that a given individual feels there is between his confidence
(or conviction) that an event $A$ will occur and his confidence (or conviction) that an event $B$
will occur. For any three events $A$, $B$ and $C$, it will be assumed that an individual is capable
of deciding whether or not the orderings $S(A,B) > S(A,C)$ and $S(A,B) < S(A,C)$ are applicable.
The notation $S(A,B) = S(A,C)$ will be used to represent the case where neither of these orderings
apply. To clarify, it is not being assumed that $S(A,B)$ and $S(A,C)$ are necessarily numerical
quantities.
Furthermore, for any fourth event $D$, it will not be assumed, in general, that an individual is
capable of deciding whether or not the orderings $S(A,B) > S(C,D)$ and $S(A,B) < S(C,D)$ are
applicable.
Therefore, a similarity $S(A,B)$ can be categorised as a partially orderable attribute of any given
pair of events $A$ and $B$. This is exactly the same definition of the concept of similarity as
used in Bowater~(2018b) and is essentially the same definition of this concept as used in
Bowater~(2017a) and Bowater~(2017b).

\vspace{3.75ex}
\noindent
\textbf{Definition 2: Discrete physical probabilities}

\vspace{1.5ex}
\noindent
Let $O=\{ O_1, O_2, \ldots, O_k \}$ be a finite ordered set of $k$ mutually exclusive, exhaustive
and equally likely outcomes of a well-understood physical experiment, which means that an outcome
is the random drawing out of a particular object from a known population of objects, e.g.\ randomly
drawing a ball out of an urn containing $k$ distinctly labelled balls.

To clarify, it will be assumed that if $O(1)$ and $O(2)$ are two subsets of the set $O$ that
contain the same number of outcomes, then the following is true:
\vspace{0.5ex}
\[
S \left( \bigcup_{O_{\hspace{-0.05em}j} \in\hspace{0.05em} O(1)} O_{\hspace{-0.05em}j},
\bigcup_{O_{\hspace{-0.05em}j} \in\hspace{0.05em} O(1)} O_{\hspace{-0.05em}j} \right)
=\, S \left( \bigcup_{O_{\hspace{-0.05em}j} \in\hspace{0.05em} O(1)} O_{\hspace{-0.05em}j},
\bigcup_{O_{\hspace{-0.05em}j} \in\hspace{0.05em} O(2)} O_{\hspace{-0.05em}j} \right)
\vspace{1.5ex}
\]
for all possible choices of the subsets $O(1)$ and $O(2)$. In making this assumption, we have
therefore, in effect, defined the circumstances in which the outcomes
$O=\{ O_1, O_2,$ $\ldots, O_k \}$ would be described as being `equally likely' to occur without
using an already established concept of probability.
Also, we have, in effect, defined what is meant by `randomly' drawing a ball out of an urn of balls
in the example that was just mentioned.

Under the assumptions that have just been made, an event $E$ that is defined by:
\begin{equation}
\label{equ12}
E = \bigcup_{O_{\hspace{-0.05em}j} \in\hspace{0.05em} O(E)} O_{\hspace{-0.05em}j}
\end{equation}
where $O(E)$ is a given subset of the set $O$, will have the probability:
\vspace{-0.5ex}
\begin{equation}
\label{equ14}
P(E) = |O(E)|\hspace{0.05em}/\hspace{0.05em}k
\vspace{-0.5ex}
\end{equation}
in which $|O(E)|$ denotes the number of outcomes in the set $O(E)$.

\vspace{4ex}
\noindent
\textbf{Definition 3: Continuous physical probabilities}

\vspace{1.5ex}
\noindent
Let $V$ be the outcome of a well-understood physical experiment that must take a value in the
interval $\Lambda=(0,1)$. Also, it will be assumed that if $\Lambda(1)$ and $\Lambda(2)$ are two
subsets of the interval $(0,1)$ that have the same total length, then the following is true:
\[
S(\{\hspace{0.05em}V\! \in \Lambda(1)\}, \{\hspace{0.05em}V\! \in \Lambda(1)\}) =
S(\{\hspace{0.05em}V\! \in \Lambda(1)\}, \{\hspace{0.05em}V\! \in \Lambda(2)\})
\]
for all possible choices of the subsets $\Lambda(1)$ and $\Lambda(2)$.

For instance, let us consider the act of randomly spinning a wheel of unit circumference, and let
us assume that any specific position on the circumference of the wheel is measured as the distance
in a given direction around the circumference from a given point on the circumference.
Here, the outcome of a spin of the wheel, as defined by the position on the circumference of the
wheel when it stops that is indicated by a fixed pointer in its centre, could be regarded as being
an example of the variable $V$.

Under the assumptions that have just been made, an event $E$ that is defined by:
\begin{equation}
\label{equ13}
E = \{\hspace{0.05em}V\! \in \Lambda(E) \}
\end{equation}
where $\Lambda(E)$ is a given subset of the interval $(0,1)$, will have the probability:
\begin{equation}
\label{equ15}
P(E) = |\Lambda(E)|
\end{equation}
in which $|\Lambda(E)|$ denotes the total length of the set $\Lambda(E)$.

\vspace{4ex}
\noindent
\textbf{Properties of physical probabilities}

\vspace{1.5ex}
\noindent
It should be clear that physical probabilities have all the standard properties of the concept of
probability, e.g.\ properties (1) to (4) in the list of such properties that was given in the
Introduction.

\vspace{3ex}
\section{Analogical probabilities}
\label{sec2}

\vspace{0.5ex}
\subsection{Reference sets of events}
\label{sec5}

\vspace{2.5ex}
\noindent
\textbf{Definition 4: Discrete reference set of events}

\vspace{1.5ex}
\noindent
Under the assumptions of Definition~2 (discrete physical probabilities), a discrete reference set
of events $R$ is defined by:
\vspace{0.5ex}
\begin{equation}
\label{equ1}
R = \{ R(\lambda): \lambda \in \Lambda \}
\vspace{0.5ex}
\end{equation}
where $R(\lambda) = O_1 \cup O_2 \cup \cdots \cup O_{\lambda k}$ and
$\Lambda = \{ 1/k,\, 2/k,\, \ldots, (k-1)/k \}$.

\vspace{4ex}
\noindent
\textbf{Definition 5: Continuous reference set of events}

\vspace{1.5ex}
\noindent
Under the assumptions of Definition~3 (continuous physical probabilities), a continuous reference
set of events $R$ is defined by equation~(\ref{equ1}), but now with the event $R(\lambda)$ defined
to be the event $\{\hspace{0.05em} V\hspace{-0.05em} < \lambda\}$ and the set $\Lambda$ defined as
in Definition~3, i.e.\ as the interval~$(0,1)$.

\pagebreak
\subsection{Non-additive analogical probabilities}
\label{sec3}

\vspace{0.5ex}
Let us consider defining the analogical probability of any given general event $E$, e.g.\ the event
of there being more than one centimetre of rain tomorrow or the event of a given US presidential
candidate being elected, as follows:
\vspace{0.5ex}
\begin{equation}
\label{equ2}
P(E)\hspace{0.1em} =\hspace{0.1em} \underset{\lambda\hspace{0.1em}
\in\hspace{0.1em} A}{\arg\max}\,\, S(E, R(\lambda))
\vspace{1ex}
\end{equation}
where $R(0)$ is an impossible event, while if $\lambda>0$, the event $R(\lambda)$ is as defined in
Def\-i\-ni\-tion~4 or Definition~5, and where $A$ is the set $\{0,\,1/k,\,\ldots, (k-1)/k,\, 1\}$
if the event $R(\lambda)$ is defined as in Definition~4 for $\lambda>0$, while $A$ is the interval
$[0,1]$ if the event $R(\lambda)$ is defined as in Definition~5 for $\lambda>0$. To clarify, the
probability $P(E)$ is the value of $\lambda \in A$ that maximises the similarity
$S(E, R(\lambda))$.
Here we could imagine gradually increasing $\lambda$ from a value of $\lambda$ for which the event
$R(\lambda)$ is considered less likely than the event $E$ until the point where the event
$R(\lambda)$ is no longer considered less likely than $R(\lambda)$. The value of $\lambda$ at this
point would be the probability of the event $E$.

This type of probability clearly has property (1) in the list of standard properties of the concept
of probability given in the Introduction, and it would be reasonable to assume that this type of
probability would always have property (2) in the list of the properties in question.
However, the scenario that has just been presented is overly idealised since, of course, the
following could be true for a value of $\lambda$ that satisfies the condition on the right-hand
side of equation~(\ref{equ2}):
\begin{equation}
\label{equ3}
S(E,R(\lambda)) < S(R(\lambda),R(\lambda))
\end{equation}
and indeed, the following may also be true:
\begin{equation}
\label{equ4}
S(E,R(\lambda)) < S(R(\lambda-a),R(\lambda))\ \ \mbox{and}\ \ S(E,R(\lambda)) <
S(R(\lambda+b),R(\lambda))
\end{equation}
where $a$ and $b$ are given positive constants and where $\lambda-a, \lambda+b \in A$.
In the situations identified by equations~(\ref{equ3}) and~(\ref{equ4}), it will often be the case
that the value of $\lambda$ that satisfies the condition on the right-hand side of
equation~(\ref{equ2}) will not be unique, and in fact, it may be very difficult to specify exactly
which values of $\lambda$ should be inside or out\-side the set of values of $\lambda$ that
satisfies this condition.
Therefore, in these circumstances, the probability $P(E)$ defined by equation~(\ref{equ2}) would be
imprecise, and if upper and lower limits were placed on this probability, then these limits
themselves may well also be imprecise.

Furthermore, if it was assumed that the event $R(\lambda)$ was defined according to Definition~4
for $\lambda>0$ and the value of $k$ was chosen to be small enough such that the value of $\lambda$
that satisfies the condition on the right-hand side of equation~(\ref{equ2}) was unique then, in
general, this value of $\lambda$, i.e.\ the probability $P(E)$, would not satisfy the following
condition:
\vspace{1.5ex}
\[
P(E) = 1 - P(\mbox{$E^{\hspace{0.02em}c}$}) = 1 -\hspace{0.05em} \underset{\lambda\hspace{0.1em}
\in\hspace{0.1em} \Lambda}{\arg\max}\,\, S(\mbox{$E^{\hspace{0.02em}c}$}, R(\lambda))
\vspace{1.5ex}
\]
i.e.\ the probabilities of the event $E$ and its complement $E^{\hspace{0.02em}c}$ would not be
additive.

It can be argued that it is convenient for probabilities to be additive since without this property
it would not be possible to place a standard probability distribution over any given uncertain
quantity. Taking this into account, a type of analogical probability that is additive will be put
forward in the next section. It will be recognised though that the kind of information concerning
what is felt about the uncertainty of events that is gained from allowing probabilities to be
non-additive is important. With regard to the concept of additive probability that will be
proposed, this information will be represented by what will be referred to as the strength of a
probability distribution.

\vspace{3ex}
\subsection{Additive analogical probabilities}
\label{sec4}

\vspace{2ex}
\noindent
\textbf{Basic properties of the type of probability to be developed}

\vspace{1.5ex}
\noindent
The type of analogical probability that will be developed in this section will be assumed to have
properties (1) to (3) in the list of standard properties of the concept of probability given in the
Introduction. However, it will not be assumed that the analogical probability of an event $B$ given
that an event $A$ has already occurred is always defined in accordance with the expression
$P(B\,|\,A) = P(A \cap B) / P(A)$, even in those cases where we would generally endorse the use of
the analogical probabilities of the event $A \cap B$ and the event $A$, i.e.\ we will not assume
that property (4) in the list of standard properties of probability being referred to always holds.

\pagebreak
\noindent
\textbf{Definition 6: Internal strength of a continuous distribution}

\vspace{1.5ex}
\noindent
Let a given continuous random variable $X$ of possibly various dimensions have two proposed
distribution functions $F_X(x)$ and $G_X(x)$. Also, we will specify the set of events
$\mathcal{F}[a]$ as follows:
\begin{equation}
\label{equ8}
\mathcal{F}[a] = \left\{ \{ X \in \mathcal{A} \}: \int_{\mathcal{A}} f_X(x) dx = a \right\}\ \ \
\mbox{for $a \in [0,1]$}
\vspace{1.5ex}
\end{equation}
where $\{ X \in \mathcal{A} \}$ is the event that $X$ lies in the set $\mathcal{A}$ and $f_X(x)$ is
the density function corresponding to $F_X(x)$, and we will specify the set $\mathcal{G}[a]$ in the
same way but with respect to the distribution function $G_X(x)$ instead of $F_X(x)$.

For a given discrete or continuous reference set of events $R$, we will now define the distribution
function $F_X(x)$ as being internally stronger than the distribution function $G_X(x)$ at the
resolution level $\lambda$, where $\lambda$ is any value in the set $\Lambda$ corresponding to the
set $R$, if
\begin{equation}
\label{equ10}
\underset{\mbox{\footnotesize \strut $A\hspace{-0.1em} \in\hspace{-0.05em}
\mathcal{F}[\lambda]$}}{\min}\, S( A, R(\lambda) ) > \underset{\mbox{\footnotesize \strut
$A\hspace{-0.1em} \in\hspace{-0.05em} \mathcal{G}[\lambda]$}}{\min}\, S( A, R(\lambda) )
\end{equation}

\vspace{5ex}
\noindent
\textbf{Definition 7: Internal strength of a discrete distribution}

\vspace{1.5ex}
\noindent
Let a given discrete random variable $X$ that can only take a value $x$ that belongs to the finite
or countable set $\{ x_1, x_2, \ldots \}$ have two proposed distribution functions $F_X(x)$ and
$G_X(x)$. Also, let the event $R^{\hspace{0.02em}*}(b_i)$ be the event
$\{\hspace{0.05em} V\hspace{-0.05em} < b_i\}$, where the random variable $V$ is as defined in
Definition~3 except that, in addition, this variable will be assumed to be independent from the
variable $X$, and where $b_i \in [0,1]$ for $i \in \{1,2,\ldots\}$.
Given these assumptions, we will furthermore specify the set of events $\mathcal{F}[a]$ as follows:
\vspace{1ex}
\begin{equation}
\label{equ5}
\mathcal{F}[a] = \left\{\, \bigcup_{i\hspace{0.03em}=1}^{\infty}\, (\mbox{$R^{\hspace{0.02em}*}$}
(b_i) \cap \{ X=x_i \})\, \left|\,\, \sum^{\infty}_{i\hspace{0.03em}=1}\, [\, b_i\hspace{-0.05em}
\in (0,1) \,] \leq 1\,\, \mbox{\Large $\wedge$}\, \sum_{i\hspace{0.03em}=1}^{\infty} b_i f_X(x_i)=a
\right. \right\}
\vspace{1.5ex}
\end{equation}
for $a \in [0,1]$, where $f_X(x)$ is the probability mass function corresponding to $F_X(x)$, and
$[\hspace{0.2em}]$ on the right-hand side of this equation denotes the indicator function, and we
will specify the set of events $\mathcal{G}[a]$ in the same way but with respect to the
distribution func\-tion $G_X(x)$ instead of $F_X(x)$.
To clarify, all events in the set $\mathcal{F}[a]$ \pagebreak would naturally be assigned a
probability of $a$ under the probability mass function $f_X(x)$ for the variable~$X$.

For a given discrete or continuous reference set of events $R$, we will now define the distribution
function $F_X(x)$ as being internally stronger than the distribution function $G_X(x)$ at the
resolution $\lambda$, where $\lambda \in \Lambda$, if the condition in equation~(\ref{equ10}) is
satisfied with respect to the definitions of the sets $\mathcal{F}[a]$ and $\mathcal{G}[a]$
currently being used.

\vspace{3ex}
One of the reasons for the first predicate in the definition of $\mathcal{F}[a]$ in
equation~(\ref{equ5}), i.e.\ the condition that at most only one value in the set $\{b_1, b_2,
\ldots\}$ is not equal to 0 or 1, is that without this predicate there would be an event in the set
$\mathcal{F}[a]$ that would be effectively equivalent to the event $R^{\hspace{0.02em}*}(a)$, in
particular it would be the event corresponding to setting $b_i=a$ $\forall\hspace{0.05em} i$.
In other words, there would be an event in this set that would have the undesirable property of
having a definition that does not depend on how the distribution function of interest $F_X(x)$ is
specified.
The practical importance of this issue will perhaps be more clearly seen when this definition of
$\mathcal{F}[a]$ is used again in Definition~9.

\vspace{4ex}
\noindent
\textbf{Definition 8: Elicitation of probability distributions}

\vspace{1.5ex}
\noindent
The elicitation process of a probability distribution function for any given continuous or discrete
random variable $X$ will be assumed to begin by the proposal of a distribution function $G_X(x)$
for this variable. It will then be naturally assumed that we try to adjust the distribution
function $G_X(x)$ so that it better represents what is known about the variable $X$.
In particular, another step in the elicitation process will be taken if, for an appropriate choice
of the resolution $\lambda$, an alternative distribution function $F_X(x)$ is judged as being,
according to the definitions just given, internally stronger than the distribution function
$G_X(x)$. If this is the case, then the distribution function $F_X(x)$ would become the current
proposed distribution function for $X$, and the same step would then be repeated until no
improvements to this distribution function can be made.

\vspace{3ex}
The rationale behind this way of formalising the elicitation process of a distribution function is
based on making the quite natural assumption that, for an appropriate choice of the resolution
$\lambda$, the adequacy of any given distribution function $G_X(x)$ as a representation of our
knowledge about the random variable $X$ can be measured by how large the similarities are in the
set $\{\hspace{0.05em} S( A, R(\lambda) )\hspace{-0.05em} :\hspace{-0.05em} A \in
\mathcal{G}[\lambda]\hspace{0.05em}\}$. Also, in attempting to take steps forward in the
elicitation process being discussed, it would seem quite reasonable to put more attention on trying
to increase the lowest similarities in this set without decreasing by too much, or at all, the
highest similarities in this set. Doing this would, of course, have the effect of increasing the
minimum similarity on the right-hand side of equation~(\ref{equ10}). Therefore, we have hopefully
justified the role of what has been defined as the internal strength of a distribution function in
the elicitation process in question.

However, observe that there is no guarantee that the distribution function that is elicited for
any given variable of interest will be unique, and therefore no guarantee that the probability that
is elicited for any given event or hypothesis will be unique, which is of course a property that is
also possessed by the type of probability that was discussed in Section~\ref{sec3}.
This is because there may be a set $F^{*}$ of possible distribution functions $F_X(x)$ for a given
variable $X$, each member of which is regarded to be internally stronger than any function $F_X(x)$
not in this set, but not internally stronger than any other function $F_X(x)$ within this set.
It would be hoped, though, that usually the distribution functions in the set $F^{*}$ would be
fairly similar to each other. In this type of situation, it is recommendable that any statistical
analysis that requires a distribution function for $X$ as an input incorporates a sensitivity
analysis over the functions $F_X(x)$ in the set $F^{*}$.

\vspace{4ex}
\noindent
\textbf{Definition 9: External strength of a continuous or discrete distribution}

\vspace{1.5ex}
\noindent
Let two random variables $X$ and $Y$ of possibly different dimensions have, respectively,
distribution functions $F_X(x)$ and $G_Y(y)$ that have been derived using any type of procedure,
including via the use of direct elicitation or via the use of a formal or informal system of
reasoning, e.g.\ derived by applying standard properties of the concept of probability such as the
ones listed in the Introduction.
To clarify, no assumption is being made about whether the variables $X$ and $Y$ are discrete or
continuous, e.g.\ one of these variables may be continuous, while the other one may be discrete.

Also, if the variable $X$ is continuous, then let the set of events $\mathcal{F}[a]$ be as defined
in equation~(\ref{equ8}), while if this variable is discrete, then let the set $\mathcal{F}[a]$ be
as specified in equation~(\ref{equ5}). Furthermore, depending on whether the variable $Y$ is
continuous or discrete, let the set of events $\mathcal{G}[a]$ be defined as the set
$\mathcal{F}[a]$ was defined in equation~(\ref{equ8}) or~(\ref{equ5}) but with respect to the
variable $Y$ instead of the variable $X$ and the distribution function $G_Y(y)$ instead of
$F_X(x)$. Finally, we will specify the minimum similarity $\uline{S}_{\hspace{0.04em}F}$ and the
maximum similarity $\oline{S}_{\hspace{0.03em}G}$ as follows:
\vspace{1ex}
\begin{equation}
\uline{S}_{\hspace{0.04em}F} = \underset{\mbox{\footnotesize \strut $A\hspace{-0.1em}
\in\hspace{-0.05em} \mathcal{F}[\lambda]$}}{\min}\, S( A, R(\lambda) )\ \ \mbox{and}\ \
\oline{S}_{\hspace{0.03em}G} = \underset{\mbox{\footnotesize \strut $A\hspace{-0.1em}
\in\hspace{-0.05em} \mathcal{G}[\lambda]$}}{\max}\, S( A, R(\lambda) )
\vspace{1.5ex}
\end{equation}

For a given discrete or continuous reference set of events $R$, we will now define the function
$F_X(x)$ as being externally stronger than the function $G_Y(y)$ at the resolution $\lambda$, where
$\lambda \in \Lambda$, if
\begin{equation}
\vspace{0.5ex}
\label{equ7}
\underset{\mbox{\footnotesize \strut $M\hspace{-0.1em} \in \hspace{-0.05em}
\mathcal{M}_A$}}{\max}\,\, \uline{S}_{\hspace{0.04em}F}\,\, >\,
\underset{\mbox{\footnotesize \strut $M\hspace{-0.1em} \in \hspace{-0.05em}
\mathcal{M}_B$}}{\max}\hspace{0.25em} \oline{S}_{\hspace{0.03em}G}
\vspace{1.5ex}
\end{equation}
where $\mathcal{M}_A$ and $\mathcal{M}_B$ are two given sets of reasoning processes that could be
used to evaluate the minimum similarity $\uline{S}_{\hspace{0.04em}F}$ and the maximum similarity
$\oline{S}_{\hspace{0.03em}G}$, respectively, and $M\hspace{-0.15em} \in\hspace{-0.15em}
\mathcal{M}$ denotes `\hspace{0.05em}over all reasoning processes in the set
$\mathcal{M}$\hspace{0.05em}'. The term `externally \linebreak stronger' is being used here because
we are comparing distribution functions for different random variables rather than for the same
random variable which was the case in the definitions of the `internal strength' of a distribution
function.

\vspace{3ex}
It is evident that, in any particular case, the definition of external strength just presented may
depend on the choices that are made for the sets $\mathcal{M}_A$ and $\mathcal{M}_B$.
However, in many cases, this issue can be avoided to a great extent by choosing the sets
$\mathcal{M}_A$ and $\mathcal{M}_B$ to be large enough so that they arguably contain all methods of
reasoning that are relevant to evaluating the similarities concerned, meaning that the condition in
equation~(\ref{equ7}) effectively becomes simply that
$\uline{S}_{\hspace{0.04em}F} > \oline{S}_{\hspace{0.03em}G}$.
As we will see later though, sometimes useful insights may be gained by considering cases where
$\mathcal{M}_A$ and/or $\mathcal{M}_B$ exclude potentially relevant methods of reasoning for
performing the evaluations in question.

\vspace{4ex}
\noindent
\textbf{Definition 10: Comparing the representativeness of the distributions of\\ different
variables}

\vspace{1.5ex}
\noindent
A distribution function $F_X(x)$ will be regarded as better representing our knowledge about the
variable $X$ than a distribution function $G_Y(y)$ represents our knowledge about the variable $Y$
if, for an appropriate choice of the resolution $\lambda$, the function $F_X(x)$ is regarded as
being externally stronger than the function $G_Y(y)$ according to Definition~9 with the sets
$\mathcal{M}_A$ and $\mathcal{M}_B$ chosen to contain all relevant methods of reasoning for
evaluating the similarities $\uline{S}_{\hspace{0.04em}F}$ and $\oline{S}_{\hspace{0.03em}G}$.

\vspace{3ex}
In comparison to the condition in equation~(\ref{equ10}), which is the basis of the definition of
internal strength, it is naturally appealing to have the maximum similarity
$\oline{S}_{\hspace{0.03em}G}$ on the right-hand side of equation~(\ref{equ7}) instead of the
minimum similarity over the set $\{ S( A, R(\lambda) ) : A \in \mathcal{G}[\lambda] \}$, as this of
course implies that all the similarities in this latter set will be less than any similarity in the
set $\{\hspace{0.1em} S( A, R(\lambda) ) : A \in \mathcal{F}[\lambda]\hspace{0.1em} \}$.
However, it would not have been sensible to have defined the concept of internal strength such that
the maximization instead of the minimization operator appears on the right-hand side of
equation~(\ref{equ10}), since if satisfying such a strong condition had been required in order for
a step in the elicitation process described in Definition~8 to have taken place, the ease with
which such a process could develop would have been generally impeded.

\vspace{4ex}
\noindent
\textbf{Example 1 of the application of Definition 10}

\vspace{1.5ex}
\noindent
To give an example of the application of Definition~10, let us compare a uniform distribution
function $F_X(x)$ over the interval $(0,1)$ for the output $X$ of a pseudo-random number generator
that has been carefully designed to produce approximately uniform random numbers in the interval
$(0,1)$ with a distribution function $G_Y(y)$ elicited by a given doctor for the change $Y$ in
average survival time that results from the administration of an untested new drug in comparison to
a standard drug.
We will assume that the resolution $\lambda$ is some value in the interval $[0.05,0.95]$.

Let us first observe that, all the similarities in the set
$\{ S( A, R(\lambda) ) : A \in \mathcal{F}[\lambda] \}$ may well be regarded as being quite high.
This is because the event $R(\lambda)$ is the outcome of a well-understood physical experiment,
while any event in the set $\mathcal{F}[\lambda]$ may well feel like it can be almost treated as
though it is the outcome of a well-understood physical experiment.
On the other hand, the doctor's uncertainty about whether or not any given event in the set
$\mathcal{G}[\lambda]$ will occur could be regarded as depending largely on his incomplete
knowledge about highly complex biological processes in the human body.
Therefore, it could reasonably be expected that if the sets $\mathcal{M}_A$ and $\mathcal{M}_B$
contain the simple method of direct evaluation, then the doctor would consider the function
$F_X(x)$ as being externally stronger than the function $G_Y(y)$ according to Definition~9, which
can be regarded, according to Definition~10, as a formal way of expressing the view that the
function $F_X(x)$ performs better than the function $G_Y(y)$ at representing the uncertainty that
these functions are intended to represent.

\vspace{4ex}
\noindent
\textbf{Example 2 of the application of Definition 10}

\vspace{1.5ex}
\noindent
To give a second example of the application of Definition~10, let us imagine that an election for a
state governor has five candidates, and a political analyst has assigned, using a simple process of
elicitation, analogical probabilities to the events $z_1, z_2, \redots, z_5$ of each one of these
candidates winning.
In particular, let $H_Z(z)$ be the distribution function that the analyst has placed over this
exhaustive set of events.

Also, let us suppose that there are two urns that both contain 100 balls, where each ball may be
either red or blue in colour. In the first urn, the ratio of red to blue balls is entirely unknown,
i.e.\ there may be from 0 to 100 red or blue balls in the urn. By contrast, in the second urn it is
known that there are exactly 50 red balls and 50 blue balls. We will denote the outcomes of drawing
a ball out of the first urn and the second urn in this example as the random variables $X$ and $Y$,
respectively, and the distribution functions for these two variables will be denoted as $F_X(x)$
and $G_Y(y)$, respectively.

Furthermore, we will imagine that, in eliciting the distribution function $F_X(x)$, the analyst
being referred to would choose to give a probability of 0.5 to both the events of drawing out a red
ball and drawing out a blue ball from the first urn. Clearly the same probability of 0.5 would be
assigned to these events if we used the second urn in place of the first urn simply by using the
concept of discrete physical probability outlined in Definition~2. To clarify, it will be assumed
therefore that, as far as the analyst is concerned, the distribution functions $F_X(x)$ and
$G_Y(y)$ are the same.

Since both $F_X(x)$ and $G_Y(y)$ are defined with regard to only two possible events, the sets of
events $\mathcal{F}[\lambda]$ and $\mathcal{G}[\lambda]$ will each only contain two events whatever
choice is made for the value of the resolution $\lambda$. To give a simple example, in the case
where $\lambda=0.5$, the set $\mathcal{F}[\lambda]$ just contains the events of drawing out a red
ball and drawing out a blue ball from the first urn, while the set $\mathcal{G}[\lambda]$ contains
the same two events but with respect to the second urn.
Also, with relevance to the case where $\lambda=0.5$, given the ambiguity surrounding the
uncertainty about whether or not any given one of the two events in the set $\mathcal{F}[0.5]$ will
occur, it should be fairly clear why the analyst in question is likely to decide that the
similarities between the event $R(0.5)$ as specified in Definitions~4 or~5 and the events in
$\mathcal{G}[0.5]$ are higher than the similarities between the event $R(0.5)$ and the events in
$\mathcal{F}[0.5]$.
Of course, by doing this, he would be effectively deciding that the distribution function $G_Y(y)$
is externally stronger than the distribution function $F_X(x)$ at a resolution level of 0.5
according to Definition~9, assuming that the sets $\mathcal{M}_A$ and $\mathcal{M}_B$ in this
definition are allowed to contain any relevant method of reasoning for evaluating the similarities
concerned.
A similar line of reasoning can be used to justify the analyst drawing the same conclusion with
respect to other values for the resolution $\lambda$ under the assumption that $\lambda$ is not
very close to 0 or 1.

Let us now assess the nature of the distribution function that the analyst has placed over the
possible outcomes of the state governor election race, i.e.\ the distribution function $H_Z(z)$.
Given that the factors that can influence which of the five candidates is elected are likely to be
considered vague and difficult to weigh up, it should be fairly clear why the analyst is likely to
regard this distribution function $H_Z(z)$ as being externally weaker, according to Definition~9,
than the distribution function $G_Y(y)$ just discussed, with the same assumptions in place about
the sets $\mathcal{M}_A$ and $\mathcal{M}_B$ and the range of the resolution $\lambda$ as were just
made. However, it would be much less easy to predict, under the same assumptions, whether any given
political analyst in the situation of interest would decide that what he chooses to be his
distribution function $H_Z(z)$ is externally stronger, weaker or neither stronger nor weaker than
what in the current example has been defined as the distribution function $F_X(x)$.
According to Definition~10, what has just been stated, can be regarded as simply a formal way of
saying that the analyst is likely to feel that the function $H_Z(z)$ performs worse than the
function $G_Y(y)$ but perhaps not worse than the function $F_X(x)$ in representing the uncertainty
that these functions are intended to represent.

\pagebreak
\noindent
\textbf{Definition 11: Criterion for choosing the best formally derived distribution function}

\vspace{1.5ex}
\noindent
Let $F_X(x)$ and $G_X(x)$ be two proposed distribution functions for a random variable $X$ that
have been derived using two separate methods of reasoning, and let us also assume that the random
variable $Y$ in Definition~9 is equivalent to the variable $X$.
Under these assumptions, the function $F_X(x)$ will be favoured over $G_X(x)$ as being the
distribution function of $X$ if, for an appropriate choice of the resolution $\lambda$, it is
externally stronger than $G_X(x)$ according to Definition~9 with the sets $\mathcal{M}_A$ and
$\mathcal{M}_B$ chosen to contain all relevant methods of reasoning for evaluating the similarities
$\uline{S}_{\hspace{0.04em}F}$ and $\oline{S}_{\hspace{0.03em}G}$.

\vspace{3ex}
Therefore, the function $F_X(x)$ will be favoured over $G_X(x)$ as being the distribution function
of $X$ if it can be regarded, according to Definition~10, as better representing our knowledge
about the variable $X$ than the function $G_X(x)$.
An example of the application of the criterion just given that uses a definition of the concept of
external strength that is not identical but very similar to Definition~9 was presented in
Section~3.7 of Bowater~(2018b).

\vspace{4ex}
\noindent
\textbf{Definition 12: Best reasoning system for justifying the importance of a given distribution}

\vspace{1.5ex}
\noindent
Let $F_X(x)$ be a distribution function that can be derived by using two different methods of
reasoning $M_0$ and $M_1$. Also, let us assume that, in Definition~9, the random variable $Y$ is
equivalent to the variable $X$ and the distribution function $F_X(x)$ is equivalent to the
distribution function $G_Y(y)$, which naturally implies that the maximum similarity
$\oline{S}_{\hspace{0.03em}G}$ on the right-hand side of equation~(\ref{equ7}) should become the
maximum similarity $\oline{S}_{F}$.
Under these assumptions, the method of reasoning $M_0$ will be regarded as better justifying the
adequacy of $F_X(x)$ as a representation of what is known about the variable $X$ than the method of
reasoning $M_1$ if, for an appropriate choice of the resolution $\lambda$, the condition in
equation~(\ref{equ7}) holds when the set $\mathcal{M}_A$ contains only the method of reasoning
$M_0$ and the set $\mathcal{M}_B$ contains only the method of reasoning $M_1$.

\vspace{3ex}
A detailed example of the application of the definition just given will be presented in
Section~\ref{sec8}.

\vspace{4ex}
\noindent
\textbf{Sensitivity to the choice of the resolution $\lambda$}

\vspace{1.5ex}
\noindent
A criticism that could be made of the definitions of internal and external strength that have been
set out in the present section, i.e.\ Definitions~6, 7 and 9, is that the conditions in
equations~(\ref{equ10}) and~(\ref{equ7}) on which these definitions are based may be affected by
the choice of the resolution level $\lambda$.

With regard to this issue, it is known that people generally have difficulty in weighing up the
uncertainty associated with the occurrence of events that are very unlikely or very likely to
occur, which is a disadvantage that could apply if $\lambda$ was less than say 0.05 or greater than
say 0.95.
On the other hand, it could be argued that the further that $\lambda$ is away from the value 0.5,
the greater the detail in which the characteristics of the distribution functions involved in the
Definitions~6 to~12 may be explored.

In conclusion, we will not try to pretend that inconsistencies can never arise due to the
conditions in equations~(\ref{equ10}) and~(\ref{equ7}) being satisfied for one choice of $\lambda$
but not for another choice of $\lambda$.
Nonetheless, it would be expected that, in many applications, the definitions that are based on
these conditions, i.e.\ Definitions~6, 7 and 9, will be largely insensitive to the choice made for
the value of the resolution $\lambda$ as long as we assume that $\lambda$ is not too close to 0 or
1, e.g.\ it is in the range $[0.05, 0.95]$, which is what we have assumed in the examples that have
been considered so far.

\vspace{3ex}
\subsection{Semi-additive analogical probabilities}
\label{sec6}

\vspace{0.5ex}
In the previous section, it was, in effect, assumed that analogical probabilities are fully
additive in the sense that the probability of any given union of disjoint events is obtained by
simply adding together the probabilities of the individual events concerned.
However, it may often be convenient to assume that analogical probabilities are only semi-additive
in the sense that the probabilities of some unions of disjoint events are indeed obtained via the
summation of the probabilities of the events concerned, while for other unions of disjoint events
this method is not necessarily valid.

For example, let us consider the case where a joint distribution has been placed over two given
random variables $X$ and $Y$. With reference to the properties of additive analogical probabilities
given at the start of Section~\ref{sec4}, we know that the density or mass function of the variable
$X$ conditional on an observed value $y$ of the variable $Y$ is not always defined by the formula:
$p(X\,|\, Y=y) = p(X,Y=y)/p(Y=y)$, where\hspace{0.05em} $p(X,Y)$ is the joint density or mass
function of $X$ and $Y$. However, it may be the case, in certain situations, that it is not even
convenient that the marginal distribution of $X$ is defined by marginalising the joint distribution
of $X$ and $Y$ with respect to $Y$. This is because, if we are only interested in expressing our
uncertainty about the variable $X$ rather than our uncertainty about both the variables $X$ and
$Y$ including their interdependence, then it may be better to assign a marginal distribution to $X$
directly rather than obtain such a distribution indirectly by using the joint distribution of $X$
and $Y$. Therefore, in this respect, we may sometimes wish to allow analogical probabilities to be
only semi-additive rather than fully additive.

\vspace{3ex}
\subsection{Frequentist probabilities}
\label{sec7}

\vspace{0.5ex}
Let us define the frequentist probability of a given event as the proportion of times that the
event occurs in the long run. This definition of probability is popular, and in fact, in recent
times, it can be regarded as the standard way of objectively trying to define the probability of
any given event. Nevertheless, it is a definition of probability with a clear defect, which is that
if a frequentist probability needs to be determined through the observation of the event of
interest in repeated trials, then we will never be able to determine this probability precisely,
and in fact, after any given number of trials we will only be able to make a statistical inference
about the probability concerned. Moreover, given that there is arguably no clear way in which such
a statistical inference about a frequentist probability could be formed in a purely objective
manner, it would seem reasonable to conclude that, in these circumstances, a frequentist
probability is in fact a type of subjective probability.

It is of interest, though, to consider how close a frequentist probability comes to being what, in
Section~\ref{sec1}, was defined as being a physical probability. Clearly, if outcomes are
repeatedly generated from a well-understood physical experiment as such an experiment was specified
in either Definition~2 or 3, then, as the number of trials increases, the proportion of times a
given event $E$ occurs, where $E$ is as defined in equation~(\ref{equ12}) or~(\ref{equ13}), will
tend in probability to the physical probability of the event $E$ as specified in either
equation~(\ref{equ14}) or~(\ref{equ15}). In this type of situation, it is therefore reasonable to
conclude that the frequentist probability of the event $E$ is equivalent in nature to the physical
probability of the event $E$.

On the other hand, if a probability is of the original type of frequentist probability that we
considered in this section, i.e.\ it is the long-run proportion of times a given outcome $E$ occurs
in repeated trials of an experiment that is effectively a black box, e.g.\ the proportion of times
a biased coin when tossed comes up heads, then let us consider the outcomes of a large number, say
a million, trials of this experiment.
In particular, observe that the proportion of times that the event $E$ occurs in these trials,
which we will denote as the proportion $\bm\hat{p}$, could be viewed, in certain circumstances, as
being the approximate physical probability of the event $E$ occurring in the next trial that will
take place. This is because we may be able to regard the next trial as being similar in nature to
taking a random draw from the outcomes of the trials that have already taken place with these
outcomes being treated as the outcomes $O_1, O_2, \ldots, O_k$ of the type of well-understood
physical experiment specified in Definition~2.

However, although we could use such a line of reasoning to argue that a frequentist probability of
the type under discussion is close in nature to being a physical probability, it is a line of
reasoning that is based on some quite important assumptions. First, it needs to be assumed that the
trials in question are independent from each other, and second, we need to suppose that the
proportion of times that the event $E$ occurs does not change over time.
Also, given these two assumptions, the proportion $\bm\hat{p}$ needs to be assumed to be
`approximately' equal to the proportion of times the event $E$ will occur in the long run which, as
has already been discussed, is a controversial assumption to make.

The need for the three assumptions just highlighted, which in any particular case are likely to be
very debatable assumptions, makes it clear that by applying the line of reasoning being referred
to, we would be, in fact, interpreting frequentist probability in terms of physical probability
using the concept of analogy. Therefore, it can be strongly argued that the type of frequentist
probability under discussion should be classified as being a given type of what, in
Section~\ref{sec2}, was defined to be analogical probability rather than as being a type of
`approximate' physical probability.

\vspace{3ex}
\subsection{Applying the concept of strength to the Bayesian\hspace{0.03em}-\hspace{0.03em}fiducial
controversy}
\label{sec8}

\vspace{0.5ex}
In this section, we will apply the concept of external strength to the controversy about whether
fiducial reasoning is of any use in circumstances where the fiducial density of a parameter of
interest is equal to a posterior density of the parameter that is derived by substituting a given
choice of the prior density of the parameter into Bayes' theorem.
We will choose to restrict our attention to the case where inferences need to be made about the
mean $\mu$ of a normal distribution that has a known variance $\sigma^2$ on the basis of a random
sample $x$ of $n$ values drawn from the distribution concerned, since it will be seen that the
issues that are explored in analysing this case are relevant to many other cases.
The type of fiducial inference that will be applied will be organic fiducial inference, which was
originally presented in Bowater~(2019) and further discussed in Bowater~(2020) and Bowater~(2021a)
before being clarified and modified in Bowater~(2021b).

Let it be assumed that very little or nothing was known about $\mu$ before the sample $x$ was
observed. In a Bayesian analysis, it would be quite conventional to try to represent this lack of
knowledge by placing a diffuse symmetric prior density over $\mu$ centred at some given value for
its median $\mu_0$.
Assuming this has been done, let the corresponding prior and posterior distribution functions be
denoted as $D(\mu)$ and $D(\mu\,|\,x)$, respectively.

Also, let the set of events $\mathcal{D}_{\mbox{\footnotesize $\mu$}}[a]$ be defined as the set
$\mathcal{F}[a]$ was defined in equation~(\ref{equ8}) but with respect to the variable $\mu$ rather
than the variable $X$, and the prior distribution function $D(\mu)$ rather than the generic
distribution function $F_X(x)$.
Similarly, we will define the set $\mathcal{D}_{\mbox{\footnotesize $\mu|x$}}[a]$ as the set
$\mathcal{F}[a]$ was defined in equation~(\ref{equ8}) but with respect to the variable $\mu$ and
the posterior distribution function $D(\mu\,|\,x)$.
Furthermore, in applying Definition~9, we will quite naturally assume that the similarities in the
set $\{ S( A, R(\lambda) ) : A \in \mathcal{D}_{\mbox{\footnotesize $\mu$}}[\lambda] \}$ are
evaluated before the data $x$ are observed and that this assessment is carried out by using the
simple method of direct evaluation.
Finally, it will be supposed that the resolution $\lambda$ is some value in the interval
$[0.05, 0.95]$.

Under these assumptions, it would be expected that the similarities in the set \linebreak
$\{ S( A, R(\lambda) ) : A \in \mathcal{D}_{\mbox{\footnotesize $\mu$}}[\lambda] \}$ would all be
regarded as being very low. In fact, we would expect that it would be difficult, if not impossible,
to find a directly elicited distribution function for any random variable in any context that could
be regarded as being externally weaker than the prior distribution function $D(\mu)$ according to
Definition~9.
This is because, apart from needing to satisfy the condition that it is diffuse and symmetric, the
choice of the prior density function for $\mu$ when there is very little or no prior information
about $\mu$ will be extremely arbitrary, implying that the definition of the events in the set
$\mathcal{D}_{\mbox{\footnotesize $\mu$}}[\lambda]$ will be just as arbitrary.
For example, if $\lambda=0.5$ then the set $\mathcal{D}_{\mbox{\footnotesize $\mu$}}[\lambda]$ will
contain the events $\{\mu < \mu_0\}$ and $\{\mu > \mu_0\}$ which clearly depend on the very
arbitrary choice of the prior median $\mu_0$.
A similar point about how the choice of the location of the prior density of $\mu$ can affect the
definition of the events in the set $\mathcal{D}_{\mbox{\footnotesize $\mu$}}[\lambda]$ can be made
with respect to other values of $\lambda$ in the range of interest, i.e.\ $[0.05, 0.95]$.
Furthermore, for all values of $\lambda$ in this range, the events in the set
$\mathcal{D}_{\mbox{\footnotesize $\mu$}}[\lambda]$ will of course also generally depend on the
arbitrary decision that needs to be made about how diffuse the prior density for $\mu$ should be
over the real line.

As was discussed in the Introduction, subjective probabilities do not need to have any of the
properties of physical probabilities. Moreover, the assumption was never made in
Sections~\ref{sec3} to~\ref{sec6} that analogical probabilities possess property (4) in the list of
standard properties of probability given in the Introduction.
Therefore, if we have subjectively elicited a pre-data distribution for a parameter $\theta$, and
if also we know the likelihood function of this parameter given the observed data, then we have no
special reason to assume that the most appropriate post-data distribution of $\theta$ is the
posterior distribution of $\theta$ obtained by substituting our pre-data distribution of $\theta$
into Bayes' theorem.
In addition, if we have, in some manner, obtained a post-distribution of the parameter $\theta$,
then there is no special reason to assume that the prior distribution of $\theta$ that is
consistent with this post-data distribution according to Bayes' theorem is the most appropriate
representation of our knowledge about $\theta$ before the data were observed.

On the other hand, in any particular scenario, we may feel justified in constructing the post-data
distribution of $\theta$ using the standard Bayesian approach that was just described if we feel
that, in the scenario concerned, good analogies can be made between what, in Section~\ref{sec1},
were defined as being physical probabilities and the prior probabilities of the parameter $\theta$
lying in given intervals of the real line, i.e.\ we may be able to justify the application of
Bayes' theorem by using the concept of analogy. However, although this type of strategy may prove
to be useful in many applications, it does not really get off the ground in solving the problem of
inference that is of current interest since we have effectively already established that, according
to Definition~9, the external strength of the prior distribution function of $\mu$ would be
relatively low.

Furthermore, if we look at this issue another way by insisting that the similarities in the set
$\{ S( A, R(\lambda) ) : A \in \mathcal{D}_{\mbox{\footnotesize $\mu|x$}}[\lambda] \}$ are,
nevertheless, evaluated after the data have been observed by using only Bayesian reasoning, then it
would seem difficult to argue that these similarities should be generally that much larger than the
similarities in the set
$\{ S( A, R(\lambda) ) : A \in \mathcal{D}_{\mbox{\footnotesize $\mu$}}[\lambda] \}$, assuming that
the conditions under which these latter similarities are evaluated are as described earlier.
To clarify, what is meant by Bayesian reasoning here is any system of reasoning that is related to
the way that Bayes' theorem updates the prior to the posterior density function of $\mu$ by
combining it with the like\-li\-hood function of $\mu$ given the data $x$.
In other words, with the same assumption in place about how the similarities in the set
$\{ S( A, R(\lambda) ) : A \in \mathcal{D}_{\mbox{\footnotesize $\mu|x$}}[\lambda] \}$ are
evaluated, it may not be easy for us to find a subjective distribution function that we would be
prepared to assign to any random variable in any context which we would regard, according to
Definition~9, as being externally weaker than the posterior distribution function $D(\mu\,|\,x)$.

Let us now observe that it would be considered common practice to try to approximate the posterior
distribution function $D(\mu\,|\,x)$ with a distribution function $C(\mu\,|\,x)$ that is the result
of using Bayes' theorem to update, on the basis of the data $x$, a prior density function of the
form $c(\mu)=\mbox{constant}$ $\forall \mu \in (-\infty, \infty)$.
Similar to how the set $\mathcal{D}_{\mbox{\footnotesize $\mu|x$}}[a]$ \linebreak was defined, let
the set\hspace{0.05em} $\mathcal{C}_{\mbox{\footnotesize $\mu|x$}}[a]$\hspace{0.05em} be defined as
the set $\mathcal{F}[a]$ was defined in equation~(\ref{equ8}) but with respect to the variable
$\mu$ and the distribution function $C(\mu\,|\,x)$.
However, we should point out, of course, that it would seem inappropriate to refer to
$C(\mu\,|\,x)$ as a posterior distribution function since it is based on a prior density function
$c(\mu)$ that does not have all the standard properties of a probability density function, in
particular it is clearly an improper density function.
Also if, in applying Definition~9, we assume that the similarities in both the set
$\{ S( A, R(\lambda) ) : A \in \mathcal{D}_{\mbox{\footnotesize $\mu|x$}}[\lambda] \}$ and the set
$\{ S( A, R(\lambda) ) : A \in \mathcal{C}_{\mbox{\footnotesize $\mu|x$}}[\lambda] \}$ are
evaluated after the data have been observed by using only Bayesian reasoning, then, since the
function $C(\mu\,|\,x)$ is being used as nothing more than an approximate form of the function
$D(\mu\,|\,x)$, the external strength of the function $C(\mu\,|\,x)$ relative to other distribution
functions is naturally inherited from $D(\mu\,|\,x)$, i.e.\ it must be roughly equivalent to the
external strength that is assigned to $D(\mu\,|\,x)$ relative to other distribution functions.

Having made these observations, we will now turn our attention to the application of organic
fiducial inference to the case of interest.
In doing this, the terminology and methodology that will be used corresponds to Bowater~(2019) and
Bowater~(2021b), nevertheless the way that organic fiducial inference will be applied to this case
is essentially equivalent to what was outlined in both Bowater~(2017b) and Bowater~(2018a).

Since the sample mean $\bar{x}$ is a sufficient statistic for $\mu$, it can naturally be assumed to
be the \emph{fiducial statistic} $Q(x)$ in this particular case.
For the sake of argument, let us also suppose that the \emph{primary random variable} (primary
r.v.) $\Gamma$ has a standard normal distribution. Making these assumptions effectively implies
that it is being assumed that the data set $x$ was generated by the following \emph{data generating
algorithm}:

\vspace{2ex}
\noindent
1) Generate a value $\gamma$ for the primary r.v.\ $\Gamma$ by randomly drawing this value from the
standard normal distribution.

\vspace{1.5ex}
\noindent
2) Determine the observed sample mean $\bar{x}$ by setting $\Gamma$ equal to $\gamma$ and
$\xoline{X}$ equal to $\bar{x}$ in the following expression:
\begin{equation}
\label{equ9}
\xoline{X} = \mu+(\sigma/\sqrt{n}\hspace{0.15em})\hspace{0.05em}\Gamma
\end{equation}
which effectively defines the distribution of the unobserved sample mean $\xoline{X}$.

\vspace{1.5ex}
\noindent
3) Given $\mu$ and $\sigma$, generate the data set $x$ from the joint density function of this data
set conditioned on the already generated value of the sample mean $\bar{x}$.

\vspace{2ex}
As it is being assumed that very little or nothing was known about $\mu$ before the data $x$ were
observed, it is quite natural to specify the \emph{global pre-data function} for $\mu$ as follows:
$\omega_G(\mu)=a$ for $\mu \in (-\infty,\infty)$, where $a>0$.
According to Principle~1 of Bowater~(2019), we now may define the fiducial density function of
$\mu$ by setting $\xoline{X}$ equal to $\bar{x}$ in equa\-tion~(\ref{equ9}) and then treating the
value $\mu$ in this equation as being a random variable, which implies that this density function
is alternatively specified by the expression:
\vspace{-0.5ex}
\begin{equation}
\label{equ11}
\mu\, |\, \sigma^2, x \sim \mbox{N}\hspace{0.05em} (\bar{x}, \sigma^2/n)
\vspace{-0.5ex}
\end{equation}
The validity of this density function as a post-data density function for $\mu$ clearly depends on
the argument that the density function of the primary r.v.\ $\Gamma$ after the data $x$ were
observed, i.e.\ the post-data density function of $\Gamma$, should be the same as the density
function of $\Gamma$ before the data $x$ were observed.
In the terminology of Bowater~(2019), this argument would be regarded as being the strong fiducial
\pagebreak argument applied to the example of interest.

We should point out that the fiducial distribution function of $\mu$ defined by
equation~(\ref{equ11}) is the same as the distribution function $C(\mu\,|\,x)$ that was referred to
earlier.
However, in applying Definition~9 to assess the external strength of this distribution function
relative to other distribution functions, it now will be assumed that fiducial reasoning is the
only type of reasoning that will be used to evaluate the similarities in the set
$\{ S( A, R(\lambda) ) : A \in \mathcal{C}_{\mbox{\footnotesize $\mu|x$}}[\lambda] \}$.
By fiducial reasoning it is meant any system of reasoning that directly attempts to justify the
fiducial argument that was just described.

To help us make the assessment of external strength in question, let us re-analyse one of the
balls-in-an-urn examples that were outlined in Bowater~(2017b).
In the example of interest, it is imagined that someone, who will be referred to as the selector,
randomly draws a ball out of an urn containing seven red balls and three blue balls and then,
without looking at the ball, hands it to an assistant. The assistant, by contrast, looks at the
ball, but in doing so, conceals it from the selector, and then places it under a cup. The selector
believes that the assistant smiled when he looked at the ball. Finally, the selector is asked to
assign a probability to the event that the ball under the cup is red.
We will assume that it is uncertain whether the assistant knew from the outset that the selector
would be asked to assign a probability to this particular event.

In this scenario, let us now imagine that, relative to other distribution functions of interest,
the selector wishes to evaluate the external strength of the Bernoulli distribution function
$B_Y(y)$ that corresponds to assigning a probability of 0.7 to the event that the ball under the
cup is red ($y=1$), and a probability of 0.3 to the event that it is blue ($y=0$).
This means that, if the set $\mathcal{B}[a]$ is defined as the set $\mathcal{F}[a]$ was defined in
equa\-tion~(\ref{equ5}) but with respect to the variable $Y$ instead of the variable $X$ and the
distribution function $B_Y(y)$ instead of $F_X(x)$, then the selector will need to evaluate the
similarities in the set $\{ S( A, R(\lambda)): A \in \mathcal{B}[\lambda] \}$, which of course will
contain only two similarities since, for any $\lambda \in (0,1)$, the set $\mathcal{B}[\lambda]$
can only contain two events.

In doing this, it will be assumed that the selector takes into account the fact that a smile by the
assistant would be information that could imply that it is less likely or more likely that the ball
under the cup is red. Therefore, his evaluation of the similarities in question must depend on his
subjective judgement regarding the meaning of the assistant's supposed smile. Nevertheless, he may
feel that, if the assistant had indeed smiled, he would not really have understood the smile's
meaning. Let it be assumed that this is indeed the case.

For this reason, let us briefly consider the scenario in which after drawing the ball out of the
urn, the selector had, without looking at the ball, placed it directly under the cup, rather than
giving the assistant an opportunity to look at the ball. In this case, the event of the ball under
the cup being red would usually be regarded as having a physical probability of 0.7.
Therefore, it would be expected that, for any given $\lambda \in [0.05,0.95]$, he would assess both
of the similarities in the set $\{\hspace{0.05em} S( A, R(\lambda))\hspace{-0.05em}
:\hspace{-0.05em} A \in \mathcal{B}[\lambda]\hspace{0.05em} \}$ as being equal to the highest
possible similarity that can exist between two events. Of course, such an assessment does not
directly apply if we switch back to the original scenario. Nevertheless, under the assumptions that
have been made, it would be expected that, in the original scenario, the selector would regard both
of the similarities in the set $\{ S( A, R(\lambda)): A \in \mathcal{B}[\lambda] \}$ as being at
least close to the highest possible similarity that can exist between two events.

Returning to the evaluation of the relative external strength of the fiducial distribution function
$C(\mu\,|\,x)$, let us now make an analogy between the uncertainty about the value $\gamma$ of the
primary r.v. $\Gamma$ after the data have been observed and the uncertainty about the colour of the
ball under the cup in the abstract scenario that has just been outlined.
In particular, given that very little or nothing was known about $\mu$ before the data $x$ were
observed, the event of observing the data should be akin to the event of the selector believing
that the assistant smiled when he looked at the ball chosen by the selector in the abstract
scenario in question, and hence the event of observing the data should have little or no effect on
the nature of the uncertainty that is felt about the value of $\gamma$.

As a result if, the post-data distribution function of $\Gamma$ is chosen to be equal to the
distribution function of $\Gamma$ before the data were observed, i.e.\ equal to a standard normal
distribution function, then it would be expected that, for any given $\lambda \in [0.05,0.95]$, the
relative external strength of this distribution function of $\Gamma$ when evaluated after the data
have been observed would be regarded as being similar to the relative external strength that would
be assigned by the selector to the distribution function $B_Y(y)$ in the abstract scenario under
discussion.
Taking into account that, after the data $x$ have been observed, there is a one-to-one mapping
between every possible value of $\Gamma$ and every possible value of the mean $\mu$, it can
therefore be argued that, when assessed after the data have been observed, the similarities in the
set $\{ S( A, R(\lambda)): A \in \mathcal{C}_{\mbox{\footnotesize $\mu|x$}}[\lambda] \}$ as defined
earlier should all be regarded, for any given $\lambda \in [0.05,0.95]$, as being equal or close to
the highest possible similarity that can exist between two events.

This conclusion could hardly be more different to the conclusion that was reached earlier when the
relative external strength of the same distribution function $C(\mu\,|\,x)$ was assessed under the
assumption that the only type of reasoning process that may be used to evaluate the similarities in
the set $\{ S( A, R(\lambda) ) : A \in \mathcal{C}_{\mbox{\footnotesize $\mu|x$}}[\lambda] \}$ is
Bayesian rather than fiducial reasoning.

To give a little more clarity, let us bring this section to an end by taking a look at how the
conclusions that have just been presented would be reflected in a natural application of
Definition~12 given in Section~\ref{sec4} to the example of interest. In applying Definition~12, it
will be assumed that the variable $X$ and the distribution function $F_X(x)$ that appear in this
definition are the mean $\mu$ and the distribution function $C(\mu\,|\,x)$, respectively.
Also, we will assume that the two different methods of reasoning $M_0$ and $M_1$ that are referred
to in Definition~12 are fiducial reasoning and Bayesian reasoning, respectively. Finally, let the
minimum similarity $\uline{S}_{\hspace{0.07em}C}$ and the maximum similarity
$\oline{S}_{\hspace{0.03em}C}$ be defined as follows:
\[
\uline{S}_{\hspace{0.07em}C} = \underset{\mbox{\footnotesize \strut $A\hspace{-0.1em}
\in\hspace{-0.05em} \mathcal{C}_{\mbox{\scriptsize $\mu|x$}}[\lambda]$}}{\min}\,
S( A, R(\lambda) )\ \ \mbox{and}\ \ \oline{S}_{\hspace{0.03em}C} =
\underset{\mbox{\footnotesize \strut $A\hspace{-0.1em} \in\hspace{-0.05em}
\mathcal{C}_{\mbox{\scriptsize $\mu|x$}}[\lambda]$}}{\max}\, S( A, R(\lambda) )
\]

Under these assumptions, it would be expected that, on the basis of all the observations made in
the preceding discussion, the following condition:
\vspace{-0.5ex}
\[
\underset{\mbox{\footnotesize \strut $M\hspace{-0.1em} \in \hspace{-0.05em}
\mathcal{M}_A$}}{\max}\,\, \uline{S}_{\hspace{0.07em}C}\,\, >\,
\underset{\mbox{\footnotesize \strut $M\hspace{-0.1em} \in \hspace{-0.05em}
\mathcal{M}_B$}}{\max}\hspace{0.25em} \oline{S}_{\hspace{0.03em}C}
\]
would be regarded as being satisfied for any given $\lambda \in [0.05,0.95]$ in the case where the
set $\mathcal{M}_A$ contains only fiducial reasoning and the set $\mathcal{M}_B$ contains only
Bayesian reasoning.
Therefore, it would be expected that fiducial reasoning would be formally regarded, according to
Definition~12, as being better than Bayesian reasoning in justifying the adequacy of the
distribution function $C(\mu\,|\,x)$ as a representation of what is known about the mean $\mu$
after the data have been observed.
This conclusion is, of course, consistent with the overall conclusions that were reached earlier in
this section.

\pagebreak
\section{Some closing remarks}
\label{sec9}

This paper set out to show that the concept of probability is best understood by dividing this
concept into two different types of probability, namely physical probability and analogical
probability. Physical probabilities, as were defined in Section~\ref{sec1}, are arguably in some
sense `objective' and possess all the standard properties of the concept of probability, but
probabilities of this type are inadequate for carrying out all the tasks that we would usually
expect could be performed by using some general notion of probability, e.g.\ probabilistic
inference about hypotheses on the basis of observed data. To carry out tasks of this latter type we
may use analogical probabilities, as were defined in Section~\ref{sec2}, which are undeniably
subjective probabilities, and which do not necessarily have all the standard mathematical
properties possessed by physical probabilities.

However, although analogical probabilities generally should be considered as being subjective
probabilities, this does not mean that an analogical probability of any given hypothesis being true
always needs to be treated as being a given individual's personal assessment of the probability
concerned.
This point relates to what was discussed in the Introduction regarding the distinction between
personal subjective probabilities and communal subjective probabilities. For example, the prior
probability a given individual assigns to the event of a fixed unknown parameter of interest lying
in a given interval would usually be appropriately classed as being a personal subjective
probability.
On the other hand, if we assume the method of fiducial inference outlined in Section~\ref{sec8} has
been applied to the example discussed in this earlier section in order to obtain a post-data
probability that the normal mean $\mu$ lies in a given interval, then this probability arguably
should be classed as being a communal subjective probability. This is because this
post-data\hspace{0.2em}/\hspace{0.2em}fiducial probability would be based on expressing a lack of
pre-data knowledge about $\mu$ in a way that is arguably universal or that, at the very least,
would be acceptable to many people.

Taking into account this observation and the discussion that was presented in Section~\ref{sec8},
it should be clear that we may often be able to calculate analogical probabilities for scientific
hypotheses being true that are close to being `objective' probabilities, which gives us a clear
response to those who criticise the general use of subjective probabilities in science.

\pagebreak
\noindent
\textbf{References}

\begin{description}

\setlength{\itemsep}{1ex}

\item[] Bernardo, J. M. and  Smith, A. F. M. (1994).\ \emph{Bayesian Theory}, Wiley, New York.

\item[] Bowater, R. J. (2017a).\ A formulation of the concept of probability based on the use of
experimental devices.\ \emph{Communications in Statistics:\ Theory and Methods}, \textbf{46},
4774--4790.

\item[] Bowater, R. J. (2017b).\ A defence of subjective fiducial inference.\ \emph{AStA Advances
in Statistical Analysis}, \textbf{101}, 177--197.

\item[] Bowater, R. J. (2018a).\ Multivariate subjective fiducial inference.\ \emph{arXiv.org
(Cornell University), Statistics}, arXiv:1804.09804.

\item[] Bowater, R. J. (2018b).\ On a generalised form of subjective probability.\ \emph{arXiv.org
(Cornell University), Statistics}, arXiv:1810.10972.

\item[] Bowater, R. J. (2019).\ Organic fiducial inference.\ \emph{arXiv.org (Cornell University),
Sta\-tis\-tics}, arXiv:1901.08589.

\item[] Bowater, R. J. (2020).\ Integrated organic inference (IOI):\ a reconciliation of
statistical paradigms.\ \emph{arXiv.org (Cornell University), Statistics}, arXiv:2002.07966.

\item[] Bowater, R. J. (2021a).\ A very short guide to IOI:\ a general framework for statistical
inference summarised.\ \emph{arXiv.org (Cornell University), Statistics}, arXiv:2104.11766.

\item[] Bowater, R. J. (2021b).\ A revision to the theory of organic fiducial inference.\
\emph{arXiv.org (Cornell University), Statistics}, arXiv:2111.09279.

\item[] de Finetti, B. (1937).\ La pr\'evision:\ ses lois logiques, ses sources subjectives.\
\emph{Annales de l'Institut Henri Poincar\'e}, \textbf{7}, 1--68.\ Translated into English in 1964
as `Foresight:\ its logical laws, its subjective sources' in \emph{Studies in Subjective
Probability}, Eds.\ H. E. Kyburg and H. E. Smokler, Wiley, New York, pp.\ 93--158.

\item[] Eagle, A. (2011).\ \emph{Philosophy of Probability:\ Contemporary Readings}, Routledge,\\
London.

\item[] Fine, T. L. (1973).\ \emph{Theories of Probability:\ An Examination of Foundations},
Academic Press, New York.

\item[] Fishburn, P. C. (1986).\ The axioms of subjective probability (with discussion).\
\emph{Sta\-tis\-ti\-cal Science}, \textbf{1}, 335--358.

\item[] Gillies, D. (2000).\ \emph{Philosophical Theories of Probability}, Routledge, London.

\item[] Jaynes, E. T. (2003).\ \emph{Probability Theory:\ The Logic of Science}, Cambridge
University Press, Cambridge.

\item[] Ramsey, F. P. (1926).\ \emph{Truth and Probability}.\ Originally an unpublished
manuscript.\ Presented in 1964 in \emph{Studies in Subjective Probability}, Eds.\ H. E. Kyburg and
H. E. Smokler, Wiley, New York, pp.\ 61--92.

\item[] Savage, L. J. (1954).\ \emph{The Foundations of Statistics}, Wiley, New York.

\end{description}

\end{document}